\documentclass[superscriptaddress,prl,twocolumn,tightenlines,10pt,a4paper]{revtex4}%

\usepackage{bbm}
\usepackage{amsfonts}
\usepackage{amssymb}
\usepackage{graphicx}
\usepackage{subfigure}
\usepackage{savesym}
\usepackage{amsmath}
\usepackage{txfonts}
\usepackage{multirow}
\usepackage{epstopdf}
\usepackage{umoline}
\usepackage{braket}
\usepackage{pbox}

\newcommand{\spinup}{|\!\uparrow\rangle}
\newcommand{\spindown}{|\!\downarrow\rangle}
\newcommand{\ue}{\mathrm{e}}
\newcommand{\ud}{\mathrm{d}}

\begin{document}

\title{Direct observation of dynamical quantum phase transitions in an interacting many-body system}

\author{P.~Jurcevic }%$^{1,2}$
\affiliation{Institut f\"ur Quantenoptik und Quanteninformation,\\
\"Osterreichische Akademie der Wissenschaften, Technikerstr. 21A, 6020 Innsbruck,Austria}
\affiliation{Institut f\"ur Experimentalphysik, Universit\"at Innsbruck, Technikerstr. 25, 6020 Innsbruck, Austria}

\author{H. Shen}
\affiliation{Institut f\"ur Quantenoptik und Quanteninformation,\\
\"Osterreichische Akademie der Wissenschaften, Technikerstr. 21A, 6020 Innsbruck,Austria}

\author{ P.~Hauke}
\affiliation{Institut f\"ur Quantenoptik und Quanteninformation,\\
\"Osterreichische Akademie der Wissenschaften, Technikerstr. 21A, 6020 Innsbruck,Austria}
\affiliation{Institut f\"ur Theoretische Physik, Universit\"at Innsbruck, Technikerstr. 25, 6020 Innsbruck, Austria}

\author{C.~Maier}
\affiliation{Institut f\"ur Quantenoptik und Quanteninformation,\\
\"Osterreichische Akademie der Wissenschaften, Technikerstr. 21A, 6020 Innsbruck,Austria}
\affiliation{Institut f\"ur Experimentalphysik, Universit\"at Innsbruck, Technikerstr. 25, 6020 Innsbruck, Austria}

\author{T.~Brydges}
\affiliation{Institut f\"ur Quantenoptik und Quanteninformation,\\
\"Osterreichische Akademie der Wissenschaften, Technikerstr. 21A, 6020 Innsbruck,Austria}
\affiliation{Institut f\"ur Experimentalphysik, Universit\"at Innsbruck, Technikerstr. 25, 6020 Innsbruck, Austria}

\author{C.~Hempel\footnote{Present address:ARC Centre of Excellence for Engineered Quantum Systems, School of Physics, University of Sydney, NSW, 2006, Australia}}
\affiliation{Institut f\"ur Quantenoptik und Quanteninformation,\\
\"Osterreichische Akademie der Wissenschaften, Technikerstr. 21A, 6020 Innsbruck,Austria}

\author{ B.~P.~Lanyon}
\affiliation{Institut f\"ur Quantenoptik und Quanteninformation,\\
\"Osterreichische Akademie der Wissenschaften, Technikerstr. 21A, 6020 Innsbruck,Austria}
\affiliation{Institut f\"ur Experimentalphysik, Universit\"at Innsbruck, Technikerstr. 25, 6020 Innsbruck, Austria}

\author{ M.~Heyl}
\affiliation{Max-Planck-Institut f\"ur Physik komplexer Systeme, 01187 Dresden, Germany}
\affiliation{Physik Department, Technische Universit\"at M\"unchen, 85747 Garching, Germany}

\author{R.~Blatt}
\affiliation{Institut f\"ur Quantenoptik und Quanteninformation,\\
\"Osterreichische Akademie der Wissenschaften, Technikerstr. 21A, 6020 Innsbruck,Austria}
\affiliation{Institut f\"ur Experimentalphysik, Universit\"at Innsbruck, Technikerstr. 25, 6020 Innsbruck, Austria}

\author{C.~F.~Roos}
\affiliation{Institut f\"ur Quantenoptik und Quanteninformation,\\
\"Osterreichische Akademie der Wissenschaften, Technikerstr. 21A, 6020 Innsbruck,Austria}
\affiliation{Institut f\"ur Experimentalphysik, Universit\"at Innsbruck, Technikerstr. 25, 6020 Innsbruck, Austria}
	
\date{\today}

\begin{abstract}
Dynamical quantum phase transitions (DQPTs) extend the concept of phase transitions and thus universality to the non-equilibrium regime. In this letter, we investigate DQPTs in a string of ions simulating interacting transverse-field Ising models. We observe non-equilibrium dynamics induced by a quantum quench and show for strings of up to 10 ions the direct detection of DQPTs by measuring a quantity that becomes non-analytic in time in the thermodynamic limit. Moreover, we provide a link between DQPTs and the dynamics of other relevant quantities such as the magnetization, and we establish a connection between DQPTs and entanglement production.
\end{abstract}

\maketitle

%%%%%%%%%%%%% bold paragraph starts here %%%%%%%%%%%%%%%%%

Today, the equilibrium properties of quantum matter are theoretically described with remarkable success. Yet, in recent years pioneering experiments have created novel quantum states beyond this equilibrium paradigm~\cite{Eisert:2015,Langen:2015}. Thanks to this progress, it is now possible to experimentally study exotic phenomena such as many-body localization~\cite{Schreiber:2015,Smith:2016}, pre-thermalization~\cite{Gring:2012,Neyenhuis:2016}, particle-antiparticle production in the lattice Schwinger model~\cite{Martinez:2016}, and light-induced superconductivity~\cite{Fausti:2011}. Understanding general properties of such non-equilibrium quantum states provides a significant challenge, calling for new concepts that extend important principles such as universality to the non-equilibrium realm. A general approach towards this major goal is the theory of dynamical quantum phase transitions (DQPTs)~\cite{Heyl:2013}, which extends the concept of phase transitions and thus universality to the non-equilibrium regime. In this letter, we directly observe the defining real-time non-analyticities at DQPTs in a trapped-ion quantum simulator for interacting transverse-field Ising models. Moreover, we provide a link between DQPTs and the dynamics of other relevant quantities such as the magnetization, and we establish a connection between DQPTs and entanglement production. Our work advances towards experimentally characterizing nonequilibrium quantum states and their dynamics, by offering general experimental tools that can be applied also to other inherently dynamical phenomena.

%%%%%%%% figure 1 %%%%%%%%%%%%%%%%
\begin{figure}[htb]
\begin{center}
\includegraphics[width=0.5\textwidth]{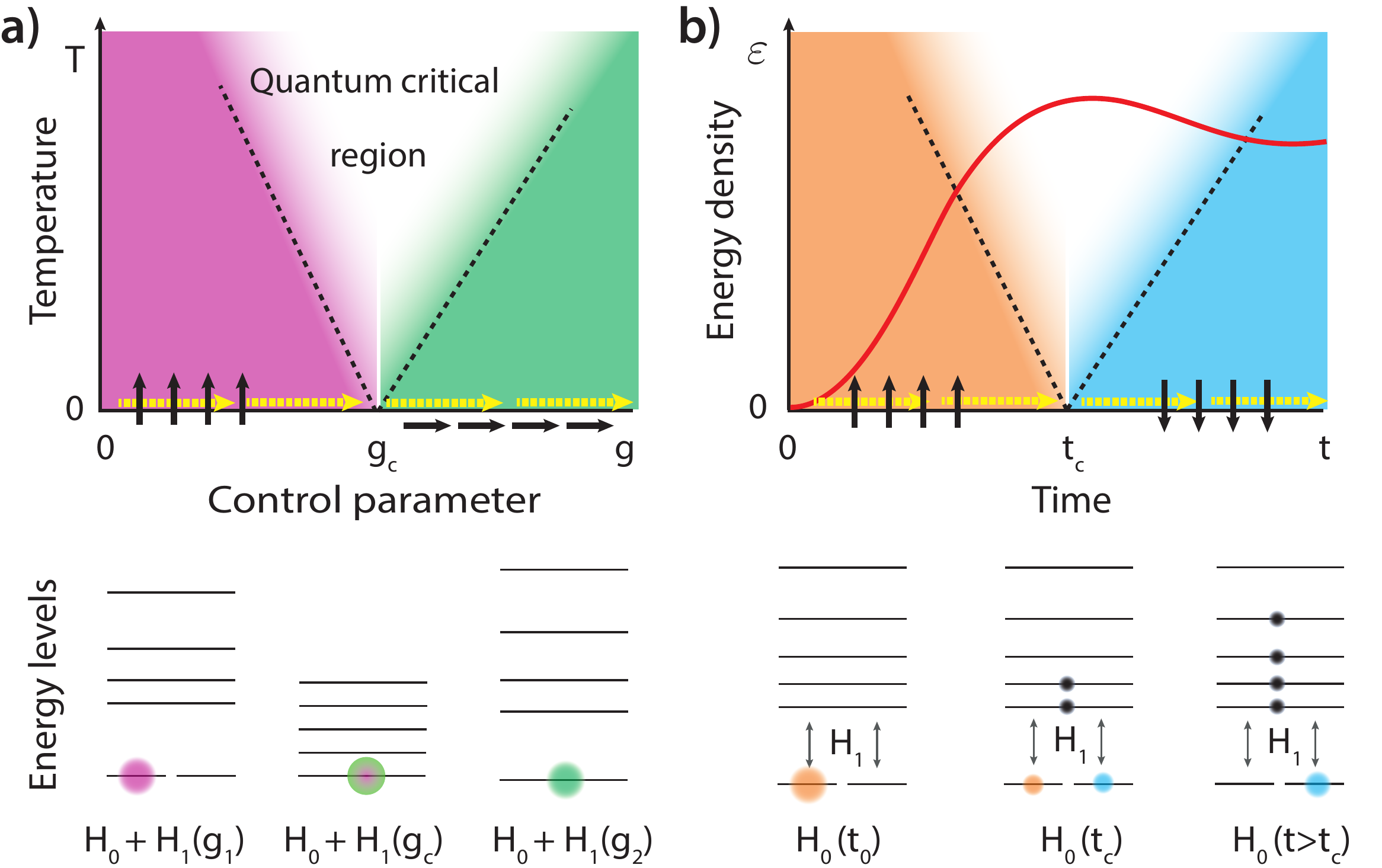}
\vspace{-4mm}
\caption{\label{fig1} \textbf{Schematic comparison of conventional and dynamical quantum phase transitions}.
\textbf{a}. 
Equilibrium temperature -- control parameter phase diagram. 
A quantum phase transition (QPT) occurs at $T=0$ separating two phases, e.g., a ferromagnet for $g<g_c$ from a paramagnet for $g>g_c$ (black arrows). 
At $g_c$, physical quantities become non-analytic upon varying $g$ (yellow arrows), triggered by a change in the spectrum of the system Hamiltonian $H=H_0+H_1(g)$, which becomes gapless as indicated by the schematic energy-level structure. 
Though only occurring at $T=0$, QPTs control the system's properties also in the quantum critical region at $T>0$. 
\textbf{b}. 
Dynamics in the energy density -- time plane.
A dynamical quantum phase transition (DQPT) occurs along the $\varepsilon=0$ axis at $t=t_c$, 
separating two regimes of, e.g., opposite magnetization (black arrows).
The DQPT is not associated with a change in the spectrum but with an incisive redistribution of occupations between the eigenstates of the initial Hamiltonian $H_0$, induced by the perturbation $H_1$. In the present experiment, $H_0$ exhibits two degenerate ground states of opposite magnetization, and the DQPT is caused by a sudden change of the low-energy occupations from one of the two ground states to the other. 
Though the mean energy density (red line), where many observables acquire their dominant contribution, lies at $\varepsilon>0$, the nonequilibrium dynamics of observables can still be controlled by the underlying DQPT (white space). 
}
\end{center}
\vspace{-5mm}
\end{figure}
%%%%%%%% figure 1 %%%%%%%%%%%%%%%%
Statistical mechanics and thermodynamics provide us with an excellent understanding of equilibrium quantum many-body systems. A key concept in this framework is the canonical partition function $Z(T)=\mathrm{Tr}(e^{-H/k_B T})$, with $T$ the temperature, $k_B$ the Boltzmann constant, and $H$ the system Hamiltonian.
The partition function encodes thermodynamics via the free-energy density $f=-(k_BT/N)\log\left[Z(T)\right]$, where $N$ denotes the number of degrees of freedom.  
A phase transition, i.e., a sudden change of macroscopic behaviour, is associated with a non-analytical behaviour of $f$ as a function of temperature or another control parameter $g$ such as an external magnetic field. 
\textit{Quantum} phase transitions (QPTs)~\cite{Sachdev2011} occur when $T$ is kept at absolute zero, where the system's ground-state properties undergo a non-analytic change as a function of $g$ (see Fig.~\ref{fig1}a). 
Scenarios where boundary conditions are essential, such as for the Casimir effect, can be studied by boundary partition functions $Z_B=\bra{\psi_0}e^{-RH}\ket{\psi_0}$, where $|\psi_0\rangle$ encodes the spatial boundary conditions on two ends of the system at distance $R$ \cite{LeClair1995}. 

Out of equilibrium, a partition function in the conventional sense cannot be formulated. Yet, remarkably, dynamical quantum phase transitions (DQPTs), where physical quantities show non-analytic behavior as a function of time, can still occur~\cite{Heyl:2013}. 
Within the theory of DQPTs, the formal role of the partition function is taken by the Loschmidt amplitude
$\mathcal{G}(t)=\bra{\psi_0}e^{-iHt}\ket{\psi_0}$, 
where $H$ is the Hamiltonian driving the time evolution and $\ket{\psi_0}$ denotes a pure quantum state, e.g., the ground state of some initial Hamiltonian $H_0$.  
Introducing a dynamical counterpart to the free-energy density, $\gamma(t)=-N^{-1}\log\left[\mathcal{G}(t)\right]$, a DQPT is defined as a non-analytic behavior in $\gamma(t)$ occuring as a function of time $t$ instead of a control parameter (see Fig.~\ref{fig1}b). 
While DQPTs represent a critical phenomenon distinct from equilibrium phase transitions (compare Fig.~\ref{fig1}a and b), many essential concepts such as universality and scaling~\cite{Heyl2015} as well as  robustness~\cite{Karrasch2013,Kriel2014,Canovi2014} carry over to them. 
Moreover, order parameters have recently been identified theoretically and also measured experimentally~\cite{Budich2016,Sharma2016,Flaeschner2016}.

Here, we report on the first direct observation of a DQPT by resolving the non-analyticity in the evolution of a quantum many-body system at a DQPT.
We demonstrate that the DQPTs are robust against modifications of microscopic details of the underlying Hamiltonian. Moreover, we provide a physical picture of how an underlying DQPT controls the dynamics of other quantities such as the magnetization and, finally, we establish a connection between the occurrence of a DQPT and entanglement production. Our work complements the very recent observation of a dynamical topological order parameter for DQPTs in systems of ultra-cold atoms in optical lattices~\cite{Flaeschner2016}.

We study DQPTs in a trapped-ion quantum simulator, realising the dynamics of an effective transverse-field Ising Hamiltonian \cite{Porras:2004,Kim:2009,Jurcevic:2014}, 
\begin{align}
H_{\rm Ising} = H_0+H_1 = -\hbar\sum_{i<j}^N J_{ij}\sigma_{i}^x\sigma^x_{j}-\hbar B\sum_i^N\sigma^z_{i}\,,
\label{eq:def_Ising}
\end{align}
with $\sigma_{i}$ Pauli spin operators on sites $i=1,\dots,N$. The coupling matrix $J_{ij}>0$ has approximately a spatial power-law dependence, $J_{ij}\sim J_{i,i+1}/|i-j|^\alpha$, with $0< \alpha < 3$ as a tunable parameter. The Hamiltonian $H_0$ exhibits spontaneous symmetry breaking with two degenerate ground states, $\ket{\Rightarrow}$ and $\ket{\Leftarrow}$, with $\sigma_i^x|\Rightarrow\rangle=|\Rightarrow\rangle$ and $\sigma_i^x|\Leftarrow\rangle=-|\Leftarrow\rangle$ $\forall i$, respectively. Recently, DQPTs in such Ising models have been studied theoretically~\cite{Zunkovic2016,Halimeh2016}.
In the experiment, the (pseudo-)spins are realized in two electronic states, e.g., 
$\ket{S_{1/2},m} \equiv \spindown_z$ and $\ket{D_{5/2}, m^\prime} \equiv \spinup_z$, of $^{40}$Ca$^+$ ions arranged in a linear string, and the encoded spins are coupled and manipulated with lasers (see Methods).

In our experiment, we adopt the following general protocol. First, the ion chain is initialized in one of the two ground states of the initial Hamiltonian $H_0$, say $\ket{\psi_0}=\ket{\Rightarrow}$. At time $t=0$, the Hamiltonian is suddenly switched to $H=H_0+H_1$, and the system state evolves to $\ket{\psi(t)}=e^{-i H t}\ket{\psi_0}$, realizing a so-called quantum quench~\cite{Eisert:2015}. At any desired time in the dynamics various observables are measured, such as the $x$-magnetization, correlation functions, or the projection onto specific states, enabling a detailed study of the DQPT.

To account for the the ground-state degeneracy of $H_0$, the Loschmidt amplitude $\mathcal{G}(t)$ is replaced by the probability to return to the ground-state manifold after a time $t$, $P(t) = P_\Rightarrow(t) + P_\Leftarrow(t)$~\cite{Heyl2014,Zunkovic2016}. 
As for $\mathcal{G}(t)$, we introduce the rate function
\begin{align}
\lambda(t) = -N^{-1}\log[P(t)]\,.
\end{align}
At the critical time $t_c$ of a DQPT, $\lambda(t)$ becomes non-analytic. 

%%%%%%%% figure 2 %%%%%%%%%%%%%%%%
\begin{figure}[t]
\begin{center}
\includegraphics[width=0.5\textwidth]{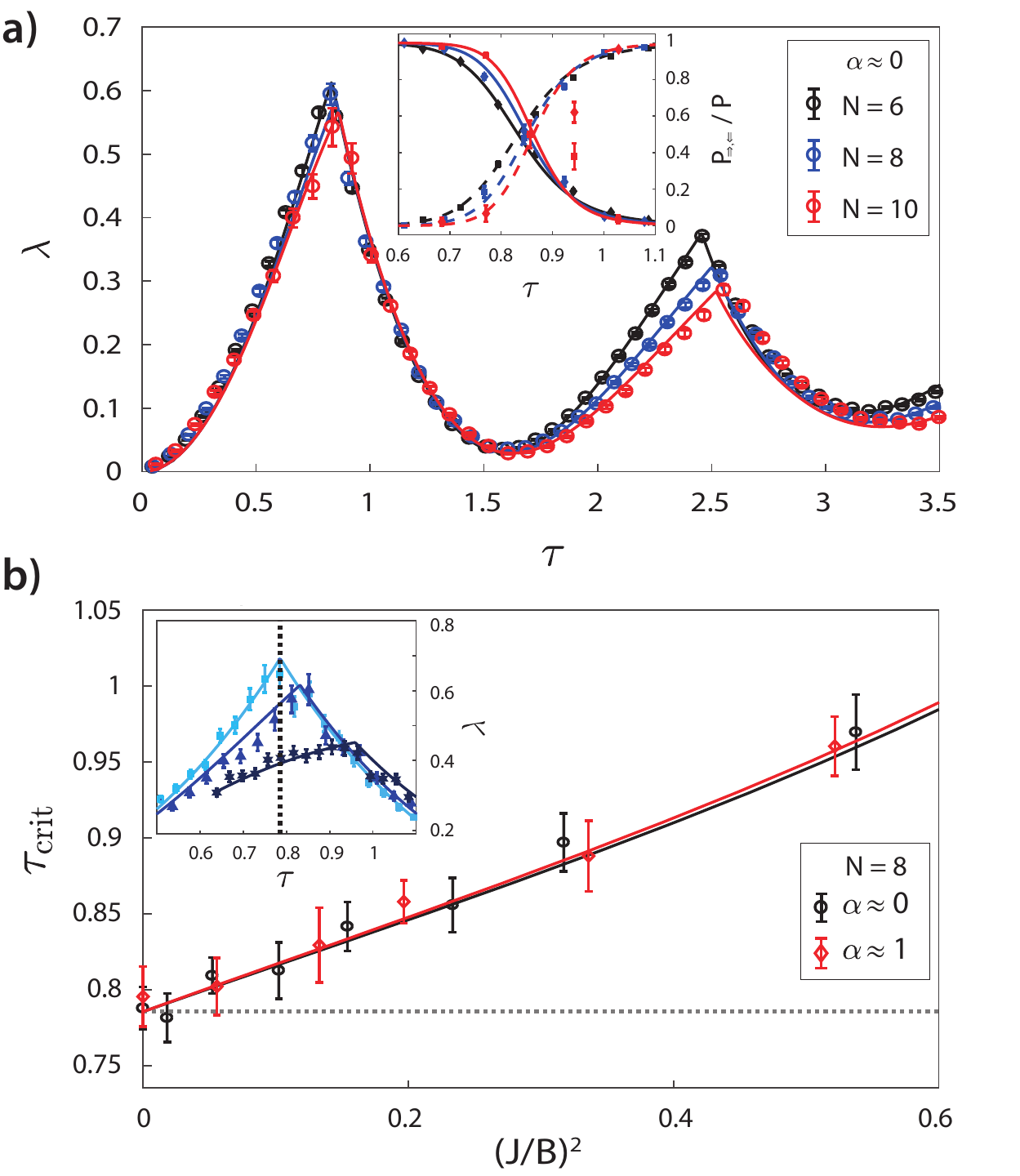}
\vspace{-4mm}
\caption{\label{fig2} \textbf{Observation of dynamical quantum phase transition.} 
\textbf{a}. Measured rate function $\lambda(\tau)$ for three different system sizes at $J/B\approx0.42$, showing a non-analytical behaviour (with $\tau=tB$ the dimensionless time).  
Dots are experimental data with error bars estimated from quantum projection noise, lines are numerical simulations with experimental parameters.
Inset: The transition between the normalized ground-state probabilities $P_{\Rightarrow, \Leftarrow}/P$ 
becomes sharper for larger $N$. 
\textbf{b}. The critical time $\tau_{\mathrm{crit}}$, i.e., the occurrence of the first DQPT, is linear as a function of $\left(J/B\right)^2$ for small $J/B$, and approximately independent of interaction range. 
\textbf{b}. The critical time $\tau_{\mathrm{crit}}$, i.e., the occurrence of the first DQPT, is linear as a function of $\left(J/B\right)^2$ for small $J/B$, and approximately independent of interaction range. 
Errorbars are $1\sigma$ confidence intervals of the fits on $\log[P_{\Rightarrow, \Leftarrow}(\tau)]$ from which we extract $\tau_{\mathrm{crit}}$ (see Methods).
Inset: DQPT exemplified for $\left(J/B\right) = 0$, $0.392$, and $0.734$. The grey dashed lines indicate $\tau_{\mathrm{crit}}$ for $\left(J/B\right)=0$.
}
\end{center}
\vspace{-5mm}
\end{figure}
%%%%%%%% figure 2 %%%%%%%%%%%%%%%%

In Figure~\ref{fig2}a, we report our first main result, the direct observation of a DQPT through non-analyticities in the rate function $\lambda$.
In our system, we can understand the origin of the measured kink by noticing that, for $N\to\infty$, $\lambda(t)$ is completely dominated by either $P_\Rightarrow(t)$ or $P_\Leftarrow(t)$, as illustrated in the inset of Fig.~\ref{fig2}a for $N\le 10$ (see also Fig.~5a). At the critical time $t=t_c$, the dominant probability switches from $P_\Rightarrow(t)$ to $P_\Leftarrow(t)$ implying that, for large $N$, $\lambda(t) = \min_{\eta\in\{\Rightarrow,\Leftarrow\}}\left(-N^{-1}\log\left[P_\eta(t)\right]\right)$~\cite{Heyl2014,Zunkovic2016}, which is the measured quantity shown in Fig.~\ref{fig2}a. We emphasize that the kinks in $\lambda$ are not caused by taking the minimum. This minimization is rather a tool to extract the DQPT for small systems without the need of extrapolating to $N\to\infty$ (see Methods and Fig.~5). The residual finite-size corrections are weak such that we can focus in the following on a single system size. 

To study the robustness of DQPTs against deformations of the Hamiltonian, we extract the first critical time $t_c$ from $\lambda(t)$ as a function of the coupling strength $J = (N-1)^{-1} \sum_{i>j} J_{ij}$, see Fig.~\ref{fig2}b. 
We find that the temporal non-analytic behavior is stable over a broad range of $J/B$ and for different $\alpha$. For $J/B \ll 1$, the critical time $\tau_c-\pi/4 \propto (J/B)^2$ exhibits a quadratic dependence on $J/B$ yielding $\tau_c= \pi/4$ for $J=0$ (see also Methods).

\begin{figure}[t]
\begin{center}
\includegraphics[width=0.5\textwidth]{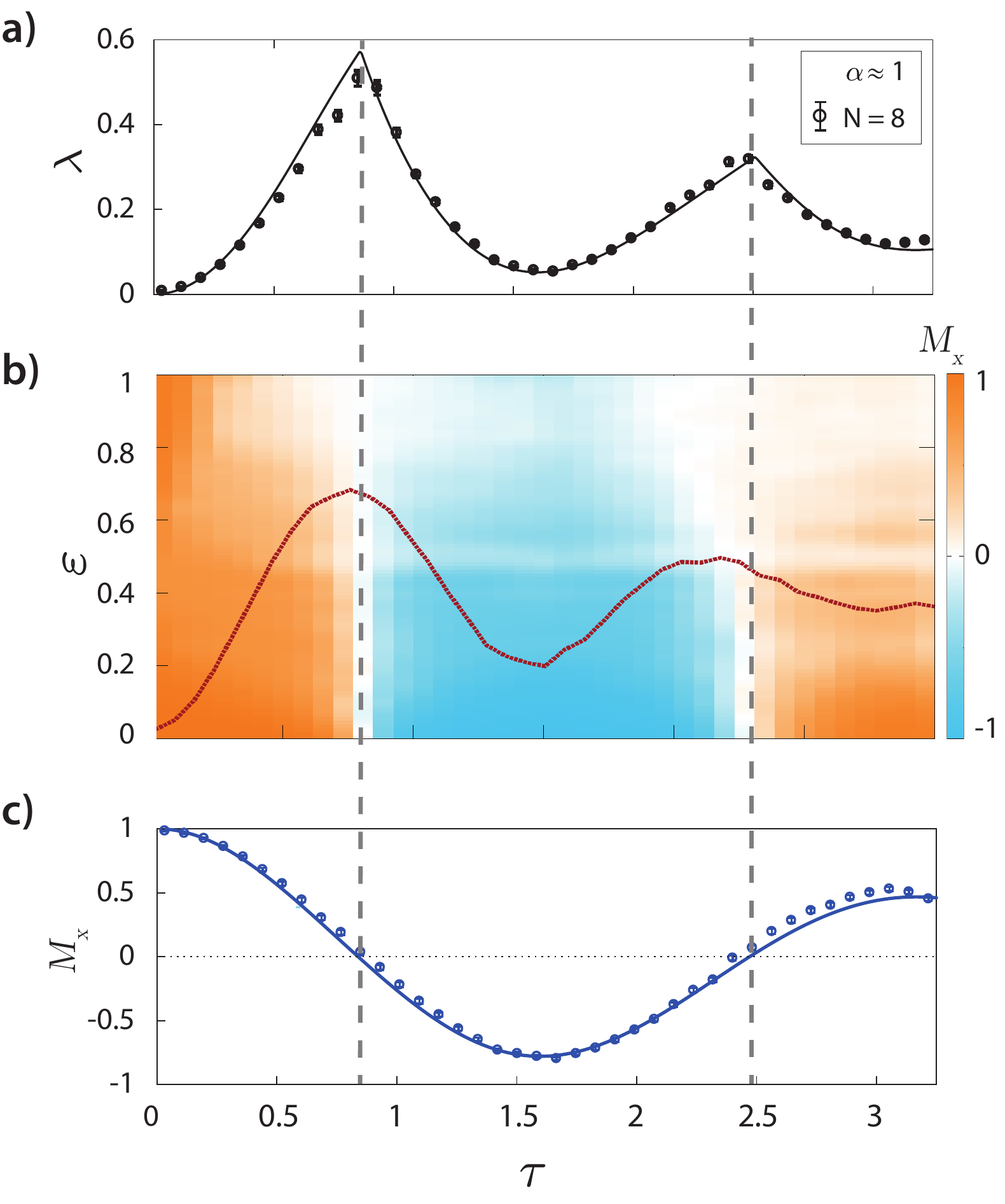}
\vspace{-4mm}
\caption{\label{fig3} \textbf{Control of the magnetisation dynamics by a DQPT}.  
DQPTs, indicated by kinks in $\lambda(\tau)$ (a), control the average magnetization in $x$-direction, $M_x$ (c). 
(b) This connection becomes apparent when resolving the magnetization against energy density $\epsilon$, with the non-analyticity at $\epsilon=0$ radiating out to $\epsilon>0$.
(b) This connection becomes apparent when resolving the magnetization against energy density $\epsilon$, with the non-analyticity at $\epsilon=0$ radiating out to $\epsilon>0$. 
For details on the measurement of the energy-resolved magnetization, see Methods. 
In (a)+(c), dots indicate experimental data with errors derived from quantum projection noise, solid lines denote numerical simulations ($J/B=0.5$).  
In (a)+(c), dots indicate experimental data with errors derived from quantum projection noise, solid lines denote numerical simulations$(J/B)=0.5$.
}
\end{center}
\vspace{-5mm}
\end{figure}

We now present measurements that connect DQPTs to other observables, further corroborating the theory of DQPT as a key framework for understanding quantum many-body dynamics. In Figs.~\ref{fig3}a and \ref{fig3}c, we compare $\lambda(t)$ and the evolution of the magnetization, $M_x(t) =  \langle \mathcal{M}_x(t)\rangle$ with $\mathcal{M}_x=N^{-1}\sum_i \sigma_i^x$. 
The initial state breaks the global $\mathbb{Z}_2$ symmetry $\sigma_i^x \to -\sigma_i^x \,\, \forall i$ of the Hamiltonian $H$. The system responds to this symmetry breaking by a repeated crossover between the $M_x>0$ and $M_x<0$ sectors, reaching the symmetry-restoring value $M_x=0$ at specific times.
Comparing with $\lambda(t)$, these are tied to the critical times of the DQPT, whose essence is the symmetry-restoration in the ground-state manifold. 

This connection is tightened by resolving the magnetization ${M}_x(\varepsilon,t)$ as a function of energy density $\varepsilon$ (see Methods and Ref.~\cite{Heyl2014}), where $\varepsilon=E/N$ and $E$ is the energy measured with the initial Hamiltonian $H_0$. The measured data is displayed in Fig.~\ref{fig3}b. 
The dynamics along $\varepsilon=0$ (ground-state manifold) is directly understood from the previous discussion. In large systems, as long as $t<t_c$ one has $P(t)\approx P_\Rightarrow(t)$, yielding ${M}_x(\varepsilon=0,t<t_c) \approx 1$. 
For $t>t_c$, $P_\Leftarrow(t)$ takes over, and ${M}_x(\varepsilon=0,t)$ jumps to $-1$. 
With increasing energy densities this sudden change smears out. 
Its influence, however, persists up to the system's mean energy density $\overline{\varepsilon}(t)$ (solid line in Fig.~\ref{fig3}b), where observables such as ${M}_x(t)$ acquire their dominant contribution~\cite{Heyl2014}.
In this way, as sketched in Fig.~\ref{fig1}, an extended region of the dynamics is controlled by the DQPT, reminiscent of a quantum critical region at an equilibrium QPT.

%
%%%%%%%% figure 3 %%%%%%%%%%%%%%%%
%%%%%%%% figure 3 %%%%%%%%%%%%%%%%
%
%%%%%%%% figure 4 %%%%%%%%%%%%%%%%
\begin{figure}[htb]
\begin{center}
\includegraphics[width=0.5\textwidth]{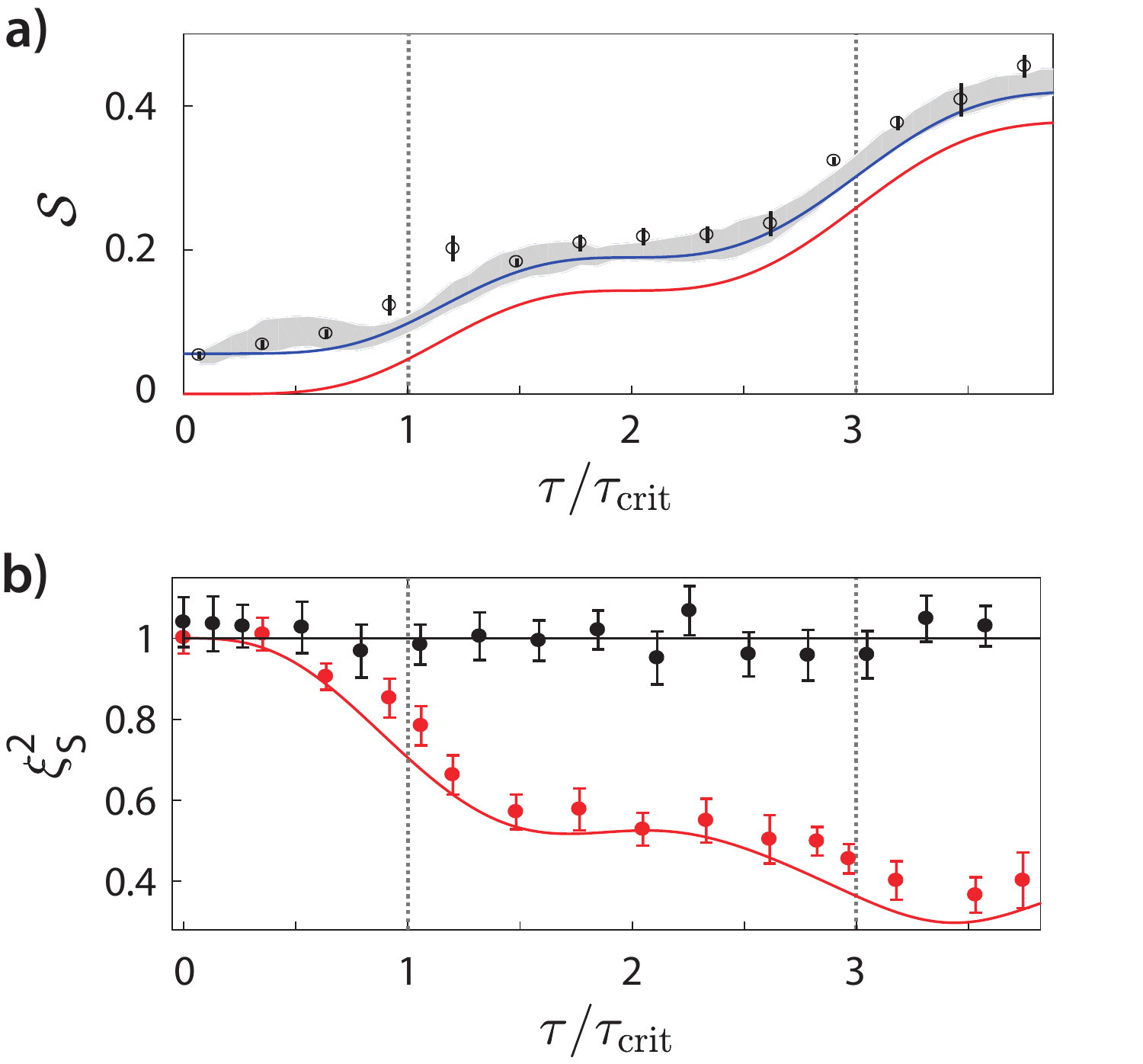}
\vspace{-4mm}
\caption{\label{fig4} \textbf{Entanglement production.} 
Dynamics of (a) the half-chain entropy $\mathcal{S}$ and (b) spin squeezing $\xi_S^2$ for $N=6$ spins at $\alpha\approx0$.
For non-zero interactions, both entanglement quantifiers show a marked increase in the vicinity of the DQPTs, indicated by dashed lines ($J/B=0.223$ in (a) and $0.25$ in (b)).
\textbf{a.} 
Comparison of the measured half-chain entropy obtained from quantum tomography (circles) with the entropies resulting from solving the Schr\"odinger equation  using our experimental parameters, with the ideal input state $\ket{\Rightarrow}$ (red line) and a slightly depolarized input state (blue line). 
Entropies obtained from simulating the tomographic reconstruction including projection noise are slightly higher, as indicated for the mixed initial state by the shaded area ($1\sigma$ confidence region). 
\textbf{b.} The change in $\xi_s^2(t)$ signals qualitatively similar entanglement production (red symbols). 
For $J/B=0$, no entanglement is created (black symbols). 
}
\end{center}
\vspace{-5mm}
\end{figure}
%%%%%%%% figure 4 %%%%%%%%%%%%%%%%

As the final result of our work, we now show that DQPTs in the simulated Ising models also control entanglement production. In this way, we connect entanglement as an important concept for the characterization of equilibrium phases and criticality~\cite{Laflorencie2016} to DQPTs.
In Fig.~\ref{fig4}a, we show the half-chain entropy  $\mathcal{S}(t)$ measured by quantum tomography (see Methods). $\mathcal{S}(t)$ exhibits its strongest growth in the vicinity of a DQPT. 
While these data are suggestive of entanglement production, $\mathcal{S}(t)$ is an entanglement measure only for pure states, which does not account for the experimentally inevitable mixing caused by decoherence. 
Therefore, we additionally measure a mixed-state entanglement witness, the Kitagawa--Ueda spin-squeezing parameter $\xi_s$ \citep{Kitagawa:1993} (see Methods) signaling entanglement whenever $\xi_s<1$. 
As Fig.~\ref{fig4}b shows, $\xi_s$ presents a behaviour qualitatively very similar to $\mathcal{S}(t)$. 
Related to common spin-squeezing scenarios \cite{Pezze2016}, 
the spin squeezing is most effective when the mean spin vector on the Bloch sphere is perpendicular to the direction of the spin-spin interaction. 
Importantly, this occurs when $M_x=0$, which we found above to be inherently tied to DQPTs. 
The presence of the DQPT, moreover, offers a more general interpretation: At exactly $t_c$, the ground-state manifold enters the equal superposition $(\ket{\Rightarrow}+\ket{\Leftarrow})/\sqrt{2}$, a highly-entangled GHZ state. 
Just as for the case of $M_x$, our data suggests that the influence of this state stretches to elevated energy densities, and thus DQPTs control also entanglement production.

We have presented the first direct observation of dynamical quantum phase transitions by revealing temporal non-analyticities in physical quantities, measured in a system of trapped ions.  
Our results provide a general approach towards experimentally accessing unifying principles of quantum many-body dynamics. Future prospects include employing the presented concepts to other nonequilibrium phenomena such as many-body localization~\cite{Schreiber:2015,Smith:2016} or quantum time crystals~\cite{Choi:2016,Zhang:2016}. 

%%%%%%%% figure 5 %%%%%%%%%%%%%%%%
\begin{figure}[t]
\begin{center}
\includegraphics[width=0.5\textwidth]{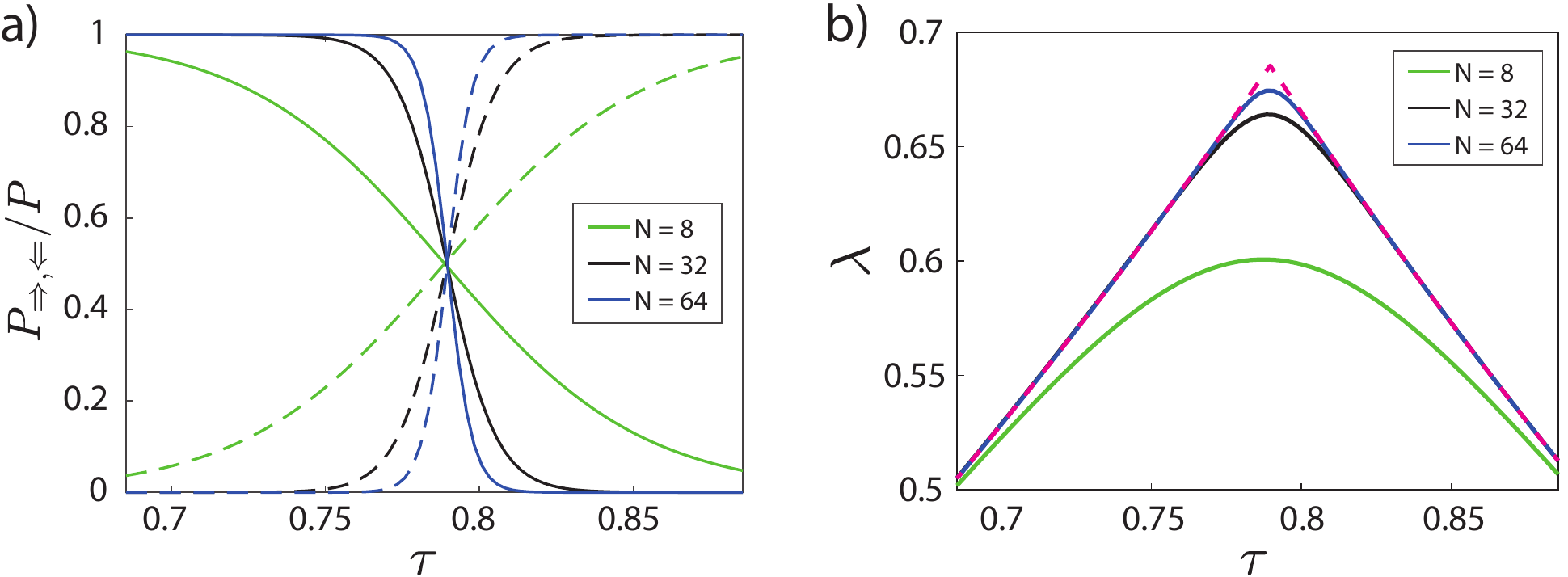}
\vspace{-4mm}
\caption{\label{extfig2}
\textbf{Scaling with system size. } 
\textbf{a.} Theory curves of the normalized probabilities $P_{\Rightarrow, \Leftarrow}/P$ for increasing numbers of spins (compare to the experimental data in the inset of Fig.~\ref{fig2}a). In the thermodynamic limit, $N \to \infty$, the transition between the two ground-state probabilities will become infinitely sharp, which is the origin of the DTQPs. 
\textbf{b.} The rate function $\lambda(\tau) = - N^{-1} \log(P(\tau))$ converges from a smooth crossover to a non-analytical kink for $N\to \infty$. The dashed line is $\min_{\eta\in\{\Rightarrow,\Leftarrow\}}\left(-N^{-1}\log\left[P_\eta(t)\right]\right)$ for $N=8$, showing that it anticipates the curve towards which $\lambda(\tau)$ converges for $N \to \infty$. 
This justifies the use of the minimum for extracting results for the thermodynamic limit in finite-size systems.  
%\ph{[in (a), can we distinguish $P_{\Rightarrow}/P$ and $P_{\Leftarrow}/P$ by solid and dashed lines?]}
%
}
\end{center}
\vspace{-5mm}
\end{figure}
%%%%%%%% figure 5 %%%%%%%%%%%%%%%%

%
\noindent {\bf Acknowledgments.} 
This work was supported by the Austrian Science Fund (FWF) under the Grant No.\ P25354-N20 and F4016-N23 (SFB FoQuS), the European Research Council (ERC) Synergy Grant UQUAM, by the Deutsche Akademie der Naturforscher Leopoldina under grant number LPDR 2015-01, by the Deutsche Forschungsgemeinschaft via the Gottfried Wilhelm Leibniz Prize programme, and by the Institut f\"ur Quanteninformation GmbH.
%

%%%%%%%%%%%%%%%%%%%%%%%%%%%%%%%%%%%%%%%%%%%%%%%%
\section{Methods}

{\bf Encoding a spin-1/2 in an optical transition of a trapped ion}.
In the experiment, $^{40}$Ca$^+$-ions are confined in a macroscopic, non-segmented, linear Paul trap.  We encode the pseudo-spins into two specific Zeeman states of the valence electron,  \mbox{$\ket{S_{1/2},m=1/2}$} and \mbox{$\ket{D_{5/2},m^\prime}$}. In order to maximize the spin-spin coupling strength in our experiments and to be able to sideband-cool the vibrational modes of interest with the available laser beams, we chose $m^\prime=5/2$ for the experiments with $\alpha=1.08$ and $m^\prime=3/2$ for the experiments with $\alpha\approx 0$.
The metastable $D_{5/2}$ state has a lifetime of 1.16(2)~s and is connected to the $S_{1/2}$ ground state by an electric quadrupole transition at a wavelength of $\lambda=729$~nm. The degeneracy of the ion's Zeeman states is lifted by a weak magnetic field of $\sim$~4~Gauss (pointing in the direction of the ion string), which allows us to initialise the \mbox{$\vert S_{1/2},m=1/2\rangle$} state using optical pumping techniques with a probability of about 99.9\% \cite{Schindler:2013}.

%%%%%%%%%%%%%%%%%
{\bf Realisation of transverse field Ising Hamiltonian with tunable spin-spin interactions. } In the following, we show how to implement the effective dynamics of the transverse-field Ising model of the main text, 
\begin{align}
H_{\rm Ising} = H_0+H_1 = -\hbar\sum_{i<j}^N J_{ij}\sigma_{i}^x\sigma^x_{j}-\hbar B\sum_i^N\sigma^z_{i}\,.
\label{eq:methods:defIsing}
\end{align}

%%%%%%%%%%%%%%%%%%%%%%%
The spin-spin coupling term $H_0$ is realized by subjecting all ions to laser pulses that off-resonantly couple the internal states to the normal modes of motion of the ions. In the limit of sufficiently weak laser fields, an effective spin-spin interaction results \cite{Porras:2004}. The spin-motion coupling used in our experiment is generated by a bichromatic beam inducing a M{\o}lmer-S{\o}rensen type interaction \cite{Sorensen:1999}. An effective transverse field described by $H_1=\hbar B\sum_i\sigma_i^z$ can be added 
by shifting both components of the bichromatic beam by an additional amount $\delta=2B$. For details of our experimental approach, see Ref.~\cite{Jurcevic:2014}. 

By laser fields coupling to all transverse motional modes, we achieve a $J_{ij}$-matrix that exhibits a spatial dependence approximately given by a power law decay $J_{ij}\propto |i-j|^{-\alpha}$ \cite{Britton:2012,Islam:2013, Nevado:2016}  which, in principle, can be tuned in the range $0< \alpha<3$. Thus, the resulting spin model cannot be mapped onto a system of non-interacting quasi-particles. The experimentally realized value of $\alpha$, which depends on the laser detuning from the normal modes and the spread of the normal mode frequencies, is inferred from the quasiparticle dispersion relation \cite{Jurcevic:2014}. 

Though the experiment realizes antiferromagnetic interactions with $J_{ij}>0$, it can still simulate the dynamics of a ferromagnetic system. The reason is that all measured observables are real, $\mathcal{O}(t)=\mathcal{O}(t)^*$, with $\mathcal{O}(t)=\braket{\psi_0|e^{i H_{\rm Ising}t}\mathcal{O}e^{-i H_{\rm Ising}t}|\psi_0}$ the time-dependent expectation value. The effect of complex conjugation is equivalent to a change of sign of the Hamiltonian, $H_{\rm{Ising}} \mapsto -H_{\rm{Ising}}$. Consequently, the dynamics of a ferromagnetic Ising model can be simulated by implementing the respective antiferromagnetic system with an opposite sign of the transverse-field $B$. 

To achieve a power-law exponent of $\alpha\approx 1.08$, experiments were carried out at axial and transverse trapping frequencies $\omega^{\mathrm{axial}}/(2\pi) = 0.219$~MHz, $\omega^{\mathrm{trans_x}}/(2\pi) = 2.72$~MHz, and $\omega^{\mathrm{trans_y}}/(2\pi) = 2.69$~MHz, and a laser detuning from the highest (the center-of-mass or `COM') mode $\Delta/(2\pi)=20$~kHz. 
In order to realize $\alpha\approx 0$, the laser fields couple to the longitudinal motional modes such that the coupling is dominated by the longitudinal COM mode. In this case, the experimental trapping parameters are $\omega^{\mathrm{axial}}/(2\pi) = 1.017$~MHz, $\omega^{\mathrm{axial}}/(2\pi) = 0.905$~MHz, and $\omega^{\mathrm{axial}}/(2\pi) = 0.743$~MHz for $N=6$, $N=8$, and $N=10$, respectively, and the detuning from the COM mode is $\Delta/(2\pi) =60$~kHz. For the six-ion experiment of Fig.~4, the resulting spin-spin coupling strength inhomogeneity is less than 5\%. 
For experiments with $\alpha=1.08$, all transverse modes are cooled to the ground state, while for $\alpha\approx 0$ only the two lowest-frequency axial modes are ground-state cooled.

%%%%%%%%%%%%%%%%%
{\bf Kac normalization of long-range interactions. }
To ensure that the contribution from the spin-spin interaction is competitive with the transverse-field term also in the long-range limit of small $\alpha$, we adopt the Kac prescription~\cite{Kac1963}. Specifically, we define a mean spin coupling $J=\sum_{i< j} J_{ij}/(N-1)$ and introduce $J/B$ as the relevant dimensionless parameter characterizing our Hamiltonian.
Using this prescription ensures the standard normalizations in the limiting cases of nearest-neighbour interactions under open boundary conditions ($J_{ij}=\delta_{j,i+1}J_0$ with $J_0=J$), as well as for infinite-range interactions ($J_{ij}=J_0$, $i\neq j$, so that $J_0=J/N$).

%%%%%%%%%%%%%%%%%
{\bf Numerical simulations of the spin dynamics. }
For numerical simulations, we use the measured trap frequencies in all three spacial dimensions and the measured Rabi-frequencies to calculate the spin-spin coupling matrix $J_{ij}$. 
The coupling matrix and the laser-ion detunings are then used to numerically calculate the time evolutions described by the effective Hamiltonian, $H_{\rm Ising}$ in Eq.~(\ref{eq:methods:defIsing}), and  acting on the initial state $\ket{\Rightarrow}$.
In the simulation, effects such as the periodic creation of virtual phonons and heating of the motional modes are disregarded. 
Moreover, we do not include any model of dephasing or depolarization, which is justified by the good agreement between theory and data over the experimental time scales. 

Due to imperfect calibrations of the Rabi frequencies and uncompensated ac-Stark shifts, the only two adjustable parameters are the mean Rabi frequency $\Omega$ and the detuning of the bichromatic light $\delta$. 
In order to find the optimal parameters describing our experimental data, we minimize the distance $D =\left( \vert P_\Rightarrow^{\mathrm{exp}} - P_\Rightarrow^{\mathrm{sim}}\vert + \vert P_\Leftarrow^{\mathrm{exp}} - P_\Leftarrow^{\mathrm{sim}}\vert + \vert M_x^{\mathrm{exp}} - M_x^{\mathrm{sim}}\vert\right)$. 
In this optimization, we strongly restrict the parameter regime, allowing $\Omega$ to be maximally $5\%$ higher than the nominal value. The inferred detuning $\delta$ differs in general by less than $\sim + 200$ Hz from the intended value.

%%%%%%%%%%%%%%%%
{\bf Measurement of quantum states and observables. }
The encoded states are discriminated by electron shelving and spatially resolved fluorescence detection on the $S_{1/2}\leftrightarrow P_{1/2}$ transition \cite{Schindler:2013}. 
In combination with arbitrary single spin rotations, we can measure the spins in any desired product basis. For example, the spins can be projected onto a combination of Pauli bases, $ \sigma_1^{\beta_1}\otimes \sigma_2^{\beta_2}\otimes \dots \otimes\sigma_N^{\beta_N}$ with $\beta = x,y,z$. 
After each projective measurement, the system is reinitialized, time-evolved, and measured again. 
This procedure is repeated between $1000$ and $5000$ times,depending on the system size and the desired observables, in order to gain sufficient statistics.
These measurements enable us to estimate the probabilities of projecting onto any of the states of the chosen measurement basis and to infer, for example, $P_{\Rightarrow}$, i.e., the probability of all spins being in the Pauli operator's eigenstate with eigenvalue +1 when measuring in the $\sigma_1^{x}\otimes \sigma_2^{x}\otimes \dots \otimes\sigma_N^{x}$ basis.
The magnetization, $M_x ={N^{-1}} \sum_{i=1}^{N}\langle\sigma_i^x\rangle/2$, is obtained by averaging the expectation values of the individual spin projections.

%%%%%%%% figure 6 %%%%%%%%%%%%%%%%
\begin{figure*}[htb]
\begin{center}
\includegraphics[width=.8\textwidth]{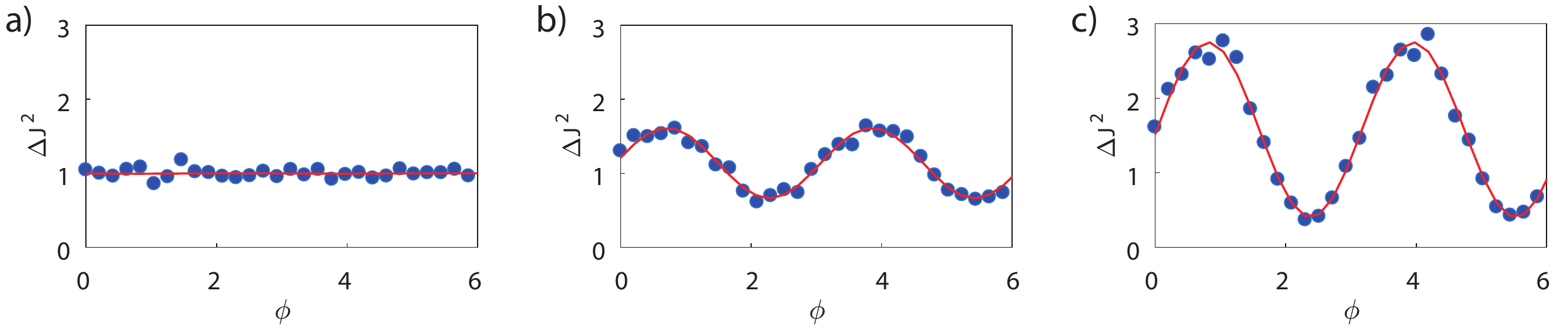}
\vspace{-4mm}
\caption{\label{extfig2}
\textbf{Spin variance $\Delta J^2$. }
The spin variance $\Delta J^2=\langle J^2\rangle-\langle J\rangle^2$ measured in the plane perpendicular to the mean spin direction $\vec{n}_0$ at different times in the dynamics: \textbf{a.} $\tau/\tau_{\mathrm{crit}} = 0$, \textbf{b.} $\tau/\tau_{\mathrm{crit}} = 1.2$ and \textbf{c.} $\tau/\tau_{\mathrm{crit}} = 3.75$. 
The experimental data (blue dots) are fitted by a sinusoidal function (red line), the minimum of which gives the minimal variance $\min\Delta J^2$.   
The $1\sigma$ confidence intervals of the corresponding fit parameters give the error bars of $\xi_s^2(t)$ as presented in Fig.~\ref{fig4}b.}  
\end{center}
\vspace{-5mm}
\end{figure*}
%%%%%%%% figure 6 %%%%%%%%%%%%%%%%

%%%%%%%%%%%%%%%%
{\bf Rate function $\lambda(t)$ and the thermodynamic limit.}
The total rate function $\lambda (t) = -N^{-1} \log\left[P(t)\right]$, with $P=P_{\Rightarrow}+P_{\Leftarrow}$, as defined in the main text, shows a non-analytical behaviour as a function of time, see Fig.~2a. 
In the thermodynamic limit, i.e., $N\rightarrow \infty$, the probability $P$ is completely dominated by either $P_{\Rightarrow}$ or $P_{\Leftarrow}$~\cite{Heyl2014,Zunkovic2016}. 
In order to show this, we express the probabilities by the rate function $\lambda_{\eta}(\tau)$,  
\begin{align}
P_{\eta} (t)= e^{-N\lambda_{\eta}(\tau)}\,,
\end{align}
with $\eta \in \{\Rightarrow, \Leftarrow\}$ for each of the two ground states. 
If at a given point in time, e.g., $\lambda_{\Rightarrow}(\tau) > \lambda_{\Leftarrow}(\tau)$, the exponential dependence of $P_\eta$ on $N$ implies $P_\Rightarrow \ll P_\Leftarrow$ leading to $P=P_\Leftarrow$ when $N \rightarrow\infty$. 
Consequently the total rate function $\lambda(\tau)$ is solely dominated by the smaller of the two $\lambda_\eta(\tau)$, yielding $\lambda(\tau)=\min_{\eta \in \{\Rightarrow, \Leftarrow\}}\lambda_\eta(\tau)$ for large $N$. 

With increasing system size $N$, the Hilbert space increases exponentially with $N$, making it experimentally challenging to estimate $P_{\Rightarrow}$ and $P_{\Leftarrow}$.
To a certain degree, we can compensate this by increasing the total number of measurements. For instance, we take 5000 measurements per data point for the ten spin case. 
For larger system sizes, the required measurement time grows as well exponentially and becomes soon infeasible. 
In order to measure the phase transitions in small-size systems, i.e., system sizes where $P_{\Rightarrow}$ and $P_{\Leftarrow}$ can be measured efficiently, we apply the knowledge that $\lambda(\tau)=\min_{\eta \in\{ \Rightarrow, \Leftarrow\}}\lambda_\eta(\tau)$ for large $N$, as shown in Fig.~5. 
As also discussed in the main text, we emphasize that this construction should be considered a tool to extract the nonanalytic behaviour already for small systems and not the origin of the non-analyticity.

%%%%%%%%%%%%%%%%
{\bf Inferring $t_{\mathrm{crit}}\,$. }
To extract $t_{\mathrm{crit}}$, we measure $\log(P_\Rightarrow(t))$ and $\log(P_\Leftarrow(t))$ in the vicinity of the phase transition, and interpolate the data points on both curves by linear fits.
Since $\log(P_\Rightarrow(t))$ and $\log(P_\Leftarrow(t))$ show a linear behaviour only over a limited range of $t$, we take for each curve the range of data points into account that gives the minimal root mean square (rms) of the linear fit. Then, $t_{\mathrm{crit}}$ is determined as the instance of time at which the fitted straight lines intersect. An errorbar for $t_{\mathrm{crit}}$ is derived from the $1\sigma$ confidence interval of the fit parameters.

%%%%%%%%%%%%%%%%%
{\bf Kitagawa-Ueda spin squeezing parameter. }
The Kitagawa-Ueda parameter $\xi_s^2$ measures spin squeezing in pure and mixed quantum states \cite{Kitagawa:1993}. To compute it, we introduce collective spin variables 
\begin{align}
J_\beta = \sum_{i=1}^N \sigma_i^\beta\,,\quad \beta = x,y,z
\end{align}
and denote the mean spin direction by $\vec{n}_0 = \langle\vec{J}\rangle/\vert \langle\vec{J}\rangle\vert$. 
The Kitagawa-Ueda spin-squeezing parameter is then defined by
\begin{align}
\xi_S^2 = 4 \frac{\min\left\lbrace\left(\Delta J_{\perp}\right)^2\right\rbrace}{N}\,,
\end{align}
where $\min\left\lbrace\left(\Delta J_{\perp}\right)^2\right\rbrace$ denotes the minimal variance perpendicular to the mean spin direction $\vec{n}_0$. If $\xi_S^2 < 1$, a state is spin squeezed, and, importantly, also exhibits quantum entanglement no matter whether the system is in a pure or mixed state \cite{Pezze2016}. 

We extract $\xi_S^2(t)$ by measuring the variance of $J_{\perp}$ as a function of $\vec{n}_\perp(\phi)$, where $\phi$ defines a rotation of the direction $\vec{n}_\perp(\phi)$ in the plane perpendicular to $\vec{n}_0$. 
The measured variances $\left(\Delta J_{\perp}\right)^2 (\phi)$ are fitted with $a+b\left(1+\sin\left(2\phi-c\right)\right)$, where $a = \min\left\lbrace\left(\Delta J_{\perp}\right)^2\right\rbrace$, see Fig.~6.
We use the $1\sigma$-confidence interval of $a$ as error bars. 
This procedure is repeated for every measured point of the time evolution.

%%%%%%%%%%%%%%%%%
{\bf Half-chain entropy.}
The half-chain entropy is a standard measure for entanglement in pure states \cite{Laflorencie2016}.
Let $\rho$ denote the density matrix representing the state of the entire system, which we partition in the middle into subsystems $A$ and $B$. The half-chain entropy $\mathcal{S}$ is defined as 
the von-Neumann entropy of the reduced density matrix $\rho_{1/2} = \mathrm{Tr}_B \, \rho$ of part A, obtained by tracing out the other half $B$, 
\begin{align}
\mathcal{S} = - \mathrm{Tr}\left( \rho_{1/2} \log\left(\rho_{1/2}\right)\right)\,.
\end{align}

For a one-dimensional system, $B$ is typically taken as the second half of the chain. However, in the special case of infinite-range interactions, $\alpha=0$, an a permutationally symmetric initial state, any equally sized bipartition can be taken as well. 
We infer the reduced half-chain density matrix by carrying out quantum state tomographies of reduced three-qubits subsystems. 
For this, we measure the reduced state in the Pauli bases $ \sigma_l^{\beta_l}\otimes \sigma_m^{\beta_m}\otimes\sigma_n^{\beta_n}$ with $\beta = x,y,z$ and $\lbrace l, m, n \rbrace= \lbrace 1,2,3\rbrace$, $\lbrace 1,2,6\rbrace$, $\lbrace 1,5,3\rbrace$, $\lbrace 1,5,6\rbrace$, $\lbrace 4,2,3\rbrace$, $\lbrace 4,2,6\rbrace$, $\lbrace 4,5,3\rbrace$ and $\lbrace 4,5,6\rbrace$. 
The choice of the measured bipartitions is given by our optimized tomography scheme that enables us to measure the 6-qubit state using a total of only $3^3=27$ different bases. 
By means of maximum-likelihood reconstruction, the reduced density matrix of any of the bipartitions given above is reconstructed from the measured expectation values. 
The errorbars are given as the $1\sigma$ distribution of all measured bipartitions.

%%%%%%%%%%%%%%%%%
{\bf Spectral decomposition.}
In the following, we describe the derivation of the spectral decomposition for the magnetization shown in Fig.~3 of the main text. We are interested in the observable 
\begin{align}
	M_x(t) = \langle \mathcal{M}_x(t) \rangle, \quad \mathcal{M}_x = \frac{1}{N} \sum_{i=1}^N \sigma_i^x \,.
\end{align}
Here, $\langle \mathcal{M}_x(t) \rangle = \langle \psi_0(t)| \mathcal{M}_x | \psi_0(t) \rangle$, with $|\psi_0(t) = e^{-iHt} | \psi_0 \rangle$ denoting the time-evolved initial state. Because the initial Hamiltonian $H_0 = -\sum_{i>j} J_{ij} \sigma_i^x \sigma_j^x$ commutes with $\mathcal{M}_x$, it is possible to find a joint eigenbasis $|E_\nu \rangle$ in which both of the operators are diagonal with corresponding real eigenvalues $E_\nu,M_\nu \in \mathbb{R}$, $H_0 |E_\nu \rangle = E_\nu |E_\nu \rangle$ and $\mathcal{M}_x |E_\nu\rangle = M_\nu | E_\nu \rangle$. 
This property allows us to write 
\begin{align}
	M_x(t) = \sum_\nu p_\nu(t) M_\nu, \quad p_\nu(t) = \Big| \langle \psi_0(t) | E_\nu \rangle \Big|^2.
\end{align}
We can introduce an equivalent continuous-energy representation of this equation. For this, we define the distribution function 
\begin{align}
	P(\varepsilon,t) = \sum_\nu p_\nu(t)\, \delta( \varepsilon - E_\nu/N ) \, ,
\end{align}
where $\varepsilon$ denotes the energy density, as well as the energy-resolved magnetization
\begin{align}
	\mathcal{M}(\varepsilon,t) = \frac{1}{P(\varepsilon,t)} \sum_\nu p_\nu(t) \, M_\nu\,  \delta( \varepsilon - E_\nu/N )\,.
\end{align}
With these two definitions, we arrive at the spectral representation for the dynamics of the mean magnetization, 
\begin{align}
	M_x(t) = \int d\varepsilon \, P(\varepsilon,t) \, \mathcal{M}(\varepsilon,t).
\end{align}
For a finite-size system, as we have in our experiment, continuous spectral quantities such as $\mathcal{M}(\varepsilon,t)$ are typically constructed by broadening the $\delta$-functions to Lorentzians, 
\begin{align}
	\delta(\varepsilon) \mapsto \delta_\mu(\varepsilon) = \frac{1}{\pi} \frac{ \mu}{\mu^2 + \varepsilon^2}\,.
\end{align}

For Fig.~3 of the main text, we have determined first the full many-body spectrum $E_\nu \in [0,W]$ of $H_0$, where $W$ is the full many-body bandwidth and where we set the ground-state energy to $E_\nu=0$ through an appropriate choice of the zero of energy. For each considered energy density $\varepsilon$, we have determined $\mathcal{M}(\varepsilon,t)$ using the  broadened $\delta$-function $\delta_\mu(\varepsilon)$ with $\mu = 1/(50 W)$. This choice for $\mu$ turned out to be the golden mean between two conditions: First, $\mu$ should be as small as possible to avoid artifacts due to the broadening. Second, $\mu$ should be sufficiently large to mitigate effects that are solely due to the discreteness of the finite-size spectrum and which will disappear at larger $N$. 

%%%%%%%%%%%%%%%%%
{\bf Perturbation theory. }
Via a perturbative expansion for the case $B\gg J$, one can gain additional quantitative insights about the system's dynamics including the magnetization and its connection to DQPTs as well as qualitative insights on the relation between DQPTs and spin squeezing. 
Going into a rotating frame with respect to $H_1$, the dynamics of an observable $\mathcal{O}$ can be written as 
\begin{equation}
	\braket{\mathcal{O}(t)} = \braket{\psi_0| U_I^\dagger \,\ue^{i H_1 t}\, \mathcal{O}\, \ue^{-i H_1 t} \,U_I |\psi_0}\,, 
\end{equation}
with $U_I   = \mathcal{T} \exp\left[-i\int_0^t \ud t' H_0^I(t') \right]$, 
where $\mathcal{T}$ denotes time ordering. The interaction Hamiltonian in the rotating frame, $H_0^I(t)=\ue^{i H_1 t} H_0 \ue^{-i H_1 t}$, reads 
\begin{eqnarray}
	H_0^I(t) &=& \sum_{i<j}J_{ij}\left[\cos^2(2Bt)\sigma_i^x \sigma_j^x + \sin^2(2Bt)\sigma_i^y \sigma_j^y \right. \\
		      & & \quad\quad \left.- 2 \sin(2Bt)\cos(2Bt)\left(\sigma_i^x \sigma_j^y+\sigma_i^y \sigma_j^x\right) \right]\,.\nonumber
		      \label{eq:Hamiltonian_I}
\end{eqnarray}
In the thermodynamic limit $N\to\infty$ for $\alpha=0$, the critical times of the DQPTs, $\tau_{\rm crit}$, coincide with the moments where the magnetization $M_x(t)$ vanishes~\cite{Zunkovic2016} (which we denote $\tau_x$ in the following). This motivates us to study the magnetization dynamics perturbatively, which will also clarify aspects of finite-size corrections for the small system sizes that we have in the experiment. Expanding $U_I$ up to second order in $J/B$ and $Jt$ using standard time-dependent perturbation theory, one obtains 
 \begin{eqnarray}
	\braket{\sigma_i^x(t)} &=& \cos(2Bt) \braket{\sigma_i^{\parallel}(t)} - \sin(2Bt) \braket{\sigma_i^{\bot}(t)}\,,\\
	\braket{\sigma_i^y(t)} &=& \sin(2Bt) \braket{\sigma_i^{\parallel}(t)} + \cos(2Bt) \braket{\sigma_i^{\bot}(t)}\,,
 \end{eqnarray}	
 with 
 \begin{eqnarray}
	\braket{\sigma_i^{\parallel}(t)} &=& 1 - C_i^{(1)}\frac{\sin^4(2Bt)}{4} - {C_i^{(2)}}\frac{4Bt-{\sin(4Bt)}}{16}\nonumber \\
	\braket{\sigma_i^{\bot}(t)} &=& -{C_i^{(1)}}\frac{8Bt-\sin(8Bt)}{64} \nonumber\\
						& & + {C_i^{(2)}}\frac{4Bt[1+2\cos(4Bt)]- \sin(4Bt)[2+\cos(4Bt)]}{16}  \,.\nonumber
 \end{eqnarray}	
Here, $\sigma_i^{\parallel}$ is the component that rotates parallel to the mean spin direction in the limit $J/B=0$, while $\sigma_i^{\bot}$ is the component in the $x-y$ plane perpendicular to it. 
The interaction strength enters via $C_i^{(1)}=\frac 1 2  \left(\sum_{j}\frac{J_{ij}}{B}\right)^2$ and $C_i^{(2)}=\frac 1 2 \sum_{j}\left(\frac{J_{ij}}{B}\right)^2$, with the mean $C^{(m)}=\frac 1 N \sum_i C_i^{(m)}$. 
For completeness, the third spin component to leading order reads $\braket{\sigma_i^z(t)} = \sum_jJ_{ij}{\sin^2(2Bt)}/{2B}$. 

At $J/B=0$, the spins perform independent Larmor precessions, and we have that $M_x=0$ precisely at the critical times $\tau_x=\tau_{\rm crit}=\pi/4$ of the DQPTs. 
At non-zero interactions $J/B>0$ both $\tau_x$ and $\tau_{\rm crit}$ obtain a shift, which for $\alpha=0$ and $N\to\infty$ is exactly identical~\cite{Zunkovic2016}. In general, however, this need not be the case, though both times remain closely connected.
Expanding $M_x(t)$ around $\pi/4$, one obtains up to second order in $J/B$, 
\begin{equation}
	\tau_{x}-\frac{\pi}{4} = \left(\frac{C^{(1)}}{2} + C^{(2)}\right)\frac{\pi}{32}\,.
\end{equation}
One can identify two different regimes. For $\alpha\leq1$, the limiting value for $N\to\infty$ is given by $C^{(1)}$ and amounts to $\tau_{x}-\frac{\pi}{4} = \frac{\pi}{32}\left(\frac J B\right)^2$ independent of $\alpha$, while $C^{(2)}$ contributes additional finite-size corrections [decreasing as $N^{-1}$ for $\alpha\leq 1/2$, as $N^{2\alpha-2}$ in the range $1/2<\alpha<1$, and as $\ln(N)^{-2}$ at $\alpha=1$].  In the second regime, $\alpha>1$, the contribution from $C^{(2)}$ converges to a non-zero value for $N\to\infty$, yielding an additional $\alpha$-dependent shift. 

Overall, the perturbative treatment for $M_x(t)$ suggests a shift of the time $\tau_x$ that depends quadratically on $J/B$. 
For the finite-size systems up to $N=10$ studied experimentally, we find that the measured shift to $\tau_x-\pi/4  \approx D_x (J/B)^2$ is different from the critical times $\tau_{\rm crit}-\pi/4\approx D_{\rm crit} (J/B)^2$ extracted from $P(t)$. Specifically, for the experimental data for $N=8$ displayed in Fig.~2b of the main text, we obtain $D_{\rm crit}\approx 0.31$ extracted from the crossing points of $P_\Rightarrow$ with $P_\Leftarrow$ and $D_x=0.18$ for the zeroes of the magnetization, which is on the same order of magnitude as the theoretical perturbative prediction of $D_x=0.097$.   
Therefore, the precise values for the finite-size critical points vary depending on the chosen observable although asymptotically they can coincide~\cite{Zunkovic2016}.

Let us close by briefly noticing that the dynamics generated by the interaction-picture Hamiltonian in Eq.~(\ref{eq:Hamiltonian_I}) provides a physical picture for the spin squeezing observed in Fig.~4b of the main text. In fact,
the dynamics in a related scenario has been successfully used to generate highly-entangled spin squeezed states \cite{Pezze2016}. 
In that scenario, one typically starts in a state polarized in the $x$-direction and applies an infinite-range interaction orthogonal to the mean spin direction, e.g., $\frac{J}{N}\sum_{i<j} \sigma_i^y \sigma_j^y$. 
In our case, due to the applied $B$ field, the spin system rotates periodically between $H_0^I(t)\propto \cos^2(2Bt)\sum_{i<j}J_{ij}\sigma_i^x \sigma_j^x$, and a squeezing interaction $H_0^I(t)\propto\sin^2(2Bt) \sum_{i<j}J_{ij}\sigma_i^y \sigma_j^y$. To lowest order, the squeezing is most efficient around $Bt=\pi/4$, i.e., at the point where the mean spin direction points on the Bloch sphere perpendicular to the interaction, and where the DQPT in the limit of $J/B\to0$ occurs. 
This dynamics acquires corrections in higher orders in $J/B$ due to the time-ordering operator $\mathcal{T}$, which leads to the shifts in time scales discussed above. 

%%%%%%%%%%%%%%%%%%%%%%%%%%%%%%%%%%%%%%%%%%%%%%%%%%%%%

%\bibliographystyle{naturemag}
%\bibliography{bibliography}

\end{document}